\documentclass[aps,prd]{revtex4}
\usepackage{subfigure,graphics} 
\usepackage{flafter} 
\begin{document} 

\title{
Brane inflation with dark reheating} 
\author{Cheng-Ji Chang}
\email{cheng-ji.chang@ncl.ac.uk}
\author{Ian G. Moss}
\email{ian.moss@ncl.ac.uk}
\affiliation{School of Mathematics and Statistics, University of  
Newcastle Upon Tyne, NE1 7RU, UK}

\date{\today}


\begin{abstract}
In brane world scenarios inflationary vacuum energy may escape into the higher
dimensional bulk leaving behind a dark radiation effect on the brane.  The
paper analyses the damping of an inflaton by massless bulk scalar radiation 
and the production of dark radiation in a Randall-Sundrum type of model.
\end{abstract}
\pacs{PACS number(s): }

\maketitle
\section{introduction}

In brane world scenarios, matter fields are confined to a four dimensional
dimensional hypersurface, or brane, embedded in a higher dimensional bulk.
Brane cosmology can reproduce the most important features of the  the more
usual four dimensional cosmology, but sometimes the extra dimensions make
themselves known and one such effect is considered here.

Many brane world scenarios are built around the Randall-Sundrum models
\cite{randall99b,randall99a}, where there is a five dimensional anti-de Sitter
bulk spacetime and either a double or a single brane.  Inflationary solutions
to the single brane  Randall-Sundrum model can be obtained by simply
`detuning' the vacuum energy of the brane \cite{Kaloper:1999sm,Nihei:1999mt}.
This can be done with the extra vacuum energy of  a brane or a bulk inflaton
field. The former case is called matter field inflation \cite{Lukas:1999yn}.

In 4-dimensional models, hot big bang radiation is produced by the decay of the
inflaton field.  In 5 dimensions, the inflaton can decay into ordinary matter
fields or into bulk fields.  In most models of inflation, the coupling between
the inflaton and other matter fields is very small, and may be smaller even
than the coupling to gravity or other bulk fields. Some of the inflaton's
vacuum energy can easily end up in the bulk. 

In the most extreme case, nearly all of the vacuum energy of the inflaton field
may decay into bulk radiation. This happens according to one scenario, put
forward by Enqvist et al. \cite{Enqvist:2003qc,enqvist04}. In their scenario,
the  entropy of the universe arises from fields associated with the flat
directions of the Minimally Supersymmetric Standard Model (MSSM). For this
scenario to work, the energy released onto the brane by these fields has to
dominate over the energy produced by the decay of the inflaton, which can
happen if the inflaton decays into bulk degrees of freedom. 

The transfer of energy from a brane to the bulk is always associated with a
geometrical effect on the brane which mimics a radiation source
term in the cosmological evolution equations
\cite{kraus99,kehagias99,cline99,binetruy99,flanagan99,hebecker01,langlois02,langlois03}. 
The size of this dark radiation term in relation to the real radiation
is an important feature, because we have strong evidence that the  real
radiation provides the dominant proportion of the radiation density driving
the expansion rate at the nucleosynthesis era.

The purpose of this paper is to construct a simple model of the inflaton decay
process with a single brane and a massless scalar bulk field. The bulk
field might occur as a hypothetical superpartner of the graviton or a bulk
dilaton field, for example. We shall construct the reheating terms in the
inflaton's equation of motion and solve for the case where bulk radiation is
the predominant form of reheating. We shall then examine the cosmological
evolution to determine the residual amount of dark radiation.

The action for the model takes the form
\begin{equation}
S_g=-{1\over \kappa_5^2}\int_{\cal M}\left (R-2\Lambda\right)\,dv
+{2\over \kappa_5^2}\int_{\cal\partial M}K\,dv
+\int_{\cal M}{\cal L}_b\,dv
+\int_{\cal \partial M}\left({\cal L}_m-\sigma\right)\,dv
\end{equation}
where we have a curved space  ${\cal M}$ with scalar curvature $R$ and a
boundary ${\cal \partial M}$ with extrinsic curvature $K$. The  covariant
volume integral is denoted by  $dv$. We will take coordinates
$x^\mu$, $\mu=0,1,2,3$,  along the brane and $x^5=z$ perpendicular to the
brane.  In the final state after inflation, the vacuum energy $\sigma$ and the
5-dimensional cosmological constant $\Lambda$ satisfy the fine tuning
relationship $6\Lambda=\kappa_5^4\sigma^2$
\cite{randall99b,randall99a}. The effective four dimensional Newton's constant
$G$ is given by $48\pi G=\kappa_5^4\sigma$.

The inflaton field $\phi(x)$ is restricted to the brane, with
Lagrangian density
\begin{equation}
\mathcal{L}_m = -\frac{1}{2}(\partial_{\mu}\phi)^{2} -V(\phi).
\label{Lphi}
\end{equation}
In the bulk, there is a massless conformally coupled scalar field $\Phi(x)$
with Lagrangian density
\begin{equation}
\mathcal{L}_b = -\frac{1}{2}(\partial_a\Phi)^{2}-\frac3{32}R\Phi^2.
\end{equation}
Conformally invariant boundary conditions fix the normal derivative of the Bulk
scalar. We consider an interaction which allows $\phi$ to spontaneously
decay into $\Phi$. Since the $\phi$ field is confined on the brane, the two
fields can only interact at $z=0$. A simple model for the Lagrangian density of
the interaction is
\begin{equation}
\mathcal{L}_{I} = -\frac{\lambda}{2}\phi(x)\Phi(x)^{2}\delta(z),
\end{equation}
where $\lambda$ is the coupling constant \cite{enqvist04}.

The effect of the radiation on the inflaton field equation is a damping term.
The general form of this term is identical to the four dimenisonal case which
can be found in ref. \cite{gleiser94}. For a flat
brane,
\begin{equation}
{d^2\phi\over dt^2}+{d V\over d\phi}=J
\end{equation}
where the leading order contribution to the dissipation is
\begin{equation}
J(t)=\lambda^{2}\textrm{Im}\int d^{4}x' G_F(x,x')^{2}
\phi(t')\theta(t-t')\Big{|}_{z=z'=0}, \label{integral}
\end{equation}
and $G_F(x,x')$ is the Feynman propagator of the bulk field $\Phi$ with Robin
boundary conditions on the brane.

For flat extra dimensions,
\begin{equation}
G_F(x,x') = \int \frac{d^{3}k}{(2\pi)^{3}}
\frac{dq}{2\pi}\,4\cos\,qz\cos\,qz' e^{i\vec{k}\cdot(\vec{x}-\vec{x}')}
\frac{i}{2\omega}e^{-i\omega(t-t')}, \label{G>}
\end{equation}
where $\omega=(\vec{k}^{2}+q^{2})^{1/2}$. The intergral (\ref{integral}) is
divergent, but we can apply dimensional regularisation. In five dimensions we
find
\begin{equation}
J(t) = -\frac{\lambda^{2}}{512\pi^{2}}\ln\left(\frac{1}{\mu}
\frac{d}{dt}\right)\left(\frac{d}{dt}\right)^{2}\phi(t), \label{resultS}
\end{equation}
where $\mu$ is the one-loop renormalisation scale. The same renormalisation
scale appears in the one-loop corrections to the potential. In order for these
corrections to remain relatively small,  $\mu$ has to be large compared to
other mass scales in potential.

The logarithmic time derivative appearing in eq. (\ref{resultS}) is a
feature of radiation back reaction effects in odd spacetime dimensions
\cite{Moss:2004dp}. For practical applications, we can use an integral kernel
to define the logarithmic derivative (see \cite{Moss:2004dp}), 
\begin{equation}
\ln\left({1\over\mu}{d\over dt}\right)\phi=
-\int_{-\infty}^t\ln\left(e^\gamma\mu(t-t')\right){d\phi\over dt'}dt'
\label{logdt}
\end{equation}
where $\gamma$ is Euler's constant.

This result which we have found applies to flat branes and flat extra
dimensions. However, we can make some simple generalisations. In the first
place, for the special case of the conformal scalar field, we can use the same
result for any bulk spacetime which is conformal to flat space. In particular,
the result also applies to anti-de Sitter extra dimensions. 

Another generalisation can be made if the brane is spatially homogeneous with
expansion rate $H$ in the range $H^2\ll \dot\phi$. In four dimensions, the
amplitude of density perturbations is of order  $H^2/\dot\phi$, and  $H^2\ll
\dot\phi$ is consistent with having small amplitude density fluctuations.  In
this case the time evolution of the inflaton is the dominant effect producing
the dissipative term in the inflaton equation and we can use the flat space
result (\ref{resultS}).

The complete form for inflaton equation of motion on an expanding brane with
$H^2\ll\dot\phi$ is
\begin{equation}
{d^2\phi\over dt^2}+3H{d\phi\over dt}+{dV\over d\phi}=
-c\ln\left({1\over \mu}{d\over dt}\right){d^2\phi\over dt^2}\label{infeq}
\end{equation}
where $c=\lambda^2/(512\pi^2)$.
The form of the damping term is strongly dependent on the fact that we have
not introduced any explicit mass terms into the bulk fields. When the same
calculation is repeated for a  bulk field of mass $M$, there is no damping of
the inflaton when $\dot\phi<M^2$. This does not, however, rule out the
possibility of higher loop or non-perturbative damping effects.

Inflation is associated with a slow roll approximation in which the
leading order terms in the inflaton equation have the smallest number of time
derivatives. The damping term has too many time  derivatives to appear at
leading order. The effects of damping will start to appear after inflation, in
the period which is usually associated with reheating. The situation is
similar to reheating, appart from the fact that, according to our assumptions,
the radiation is radiated into the bulk.

To investigate the period following inflation, consider a quadratic form of
potential with minimum at $\phi=0$,
\begin{equation}
V(\phi)=\frac12 m^2\phi^2
\end{equation}
An approximate solution can be constructed by taking
\begin{equation}
\phi(t)=A(t)\sin\theta(t)
\end{equation}
where $A(t)$ is a slowly varying function of  $t$ and $\theta\approx mt$. From
the definition (\ref{logdt}),
\begin{equation}
\ln\left({1\over \mu}{d\over dt}\right){d^2\phi\over dt^2}\approx
m^2\ln\left({m\over \mu}\right)A\sin(\theta)+m^2\left(Si(\theta)\cos(\theta)
-Ci(\theta)\sin(\theta)\right)A
\end{equation}
where $Si$ and $Ci$ are sine and cosine integrals. After solving (\ref{infeq})
for an initial value $\phi=\phi_i$ at $t=t_i$, we obtain
\begin{equation}
A(t)\approx\phi_i\left({a_i\over a}\right)^{3/2}e^{-\pi cm(t-t_i)/4}
\end{equation}
If we also set the initial condition to coincide with the end of inflation,
then $\phi_i\sim M_p$, where $M_p^2=(8\pi G)^{-1}$.

The effect of the inflaton on the expansion of the universe can be obtained
from the brane cosmology equations 
\cite{kraus99,kehagias99,cline99,binetruy99,flanagan99,shiromizu99}. 
We shall only consider the case where the intrinsic brane vacuum energy is
large compared to the density, $\sigma\ll\rho$, when
\begin{equation}
6\dot H+12 H^2=8\pi G(\rho-3p)-2\kappa_5^2T^r_{55}\label{branecos}
\end{equation}
where  $T^r_{ab}$ is the  stress-energy of the bulk radiation.  Since we are
assuming that the radiation in mostly into the extra dimensions,
the density $\rho$ and pressure $p$ are those of the inflaton field,
\begin{eqnarray}
\rho&=&\frac12\dot\phi^2+\frac12m^2\phi^2\\
p&=&\frac12\dot\phi^2-\frac12m^2\phi^2
\end{eqnarray}
The radiation term  can be obtained by similar means to the dissipative term
\cite{Moss:2004dp}. The terms up to order $\lambda^2$ are
\begin{equation}
T^r_{55}=
-{3\lambda\over 2048\pi^2}
\ln\left(\frac{1}{\mu}\frac{d}{dt}\right){d^4\phi\over dt^4}+
c'{d\phi\over dt}\ln\left(\frac{1}{\mu}\frac{d}{dt}\right){d^2\phi\over dt^2}
\end{equation}
where $c'=O(\lambda^2)$. The leading term averages out to zero over a period of
the oscillating inflaton field. 

The dark radiation $\rho_{dark}$, defined by 
\begin{equation}
3H^2=8\pi G(\rho+\rho_{dark}),\label{friedman}
\end{equation}
can be obtained by integrating (\ref{branecos}). We shall express the result in
terms of the radiation flux $T^r_{05}$, using the conservation of energy,
$T^r_{05}=-\frac12J\dot\phi$.
If we assume that
$\rho_{dark}\to 0$ as $t\to-\infty$, then
\begin{equation}
\rho_{\rm dark}=
-{1\over a^4}\int_{-\infty}^\infty\,a(t')^4\,T^r_{05}\,dt'
-{\kappa_5^2\over 8\pi G}
{1\over a^4}\int_{-\infty}^\infty H(t')a(t')^4T^r_{55}dt'
\end{equation}
The first term in identical to the amount of radiation which would be produced
if the inflaton's energy was decaying into brane radiation rather than bulk
radiation. The second term turns out to be less important than the first term
when $\sigma<\rho$.

In the early stages of reheating, the inflaton field energy density dominates
the Friedman equation but decays faster than the dark radiation density, which
will eventually come to dominate. There is some intermediate time $t_{eq}$ at
which the scalar field energy density  is equal to the dark energy density.
In the WKB approximation, after averaging over the phase, the radiation flux is
given by
\begin{equation}
T^r_{05}\approx \frac14cA^2m^3\ln\left({m\over\mu}\right)
\end{equation}
The inflaton looses energy provided that $m<\mu$. Most of the the dark energy
is generated whilst the scalar field dominates eq. (\ref{friedman}). In this
regime,
\begin{equation}
3H^2\approx4\pi Gm^2A^2
\end{equation}
We can solve this equation and evaluate the integral to obtain the dark energy
at the time $t_{eq}$,
\begin{equation}
\rho_{dark}\approx-{3\pi\over 40}c^2m^2M_p^2\ln\left({m\over\mu}\right)
\end{equation}
where $M_p^2=(8\pi G)^{-1}$. This is also the maximum value reached by the dark
radiation density.

For comparison, the value of the potential at the end of the inflationary era
would be of order $V_i\sim m^2M_p^2$. Therefore 
$\rho_{dark}/V_i\sim10^{-8}\lambda^4$. A typical inflationary model,
which was conventional in every respect appart from the reheating, might have
$V_i\sim 10^{16}$ GeV. There is no particular reason why $\lambda$ should be
very small, and a dark radiation `temperature' as high as $10^{14}$ GeV is a
possibility.

We have so far said nothing about the generation of real radiation. Let us
suppose that this is produced from a field which lies along a flat direction in
the potential. These fields have a non-vanishing potential $V_F$ as a result
of supersymmetry breaking and possible non-renormalisable terms
\cite{Enqvist:2003gh}. 
If the dark energy density $\rho_{dark}<V_F$ at $t_{eq}$, then reheating (or
more specifically preheating \cite{Kofman:1997yn}) would give a
radiation energy density $\rho_r\sim V_F$, and the ratio 
\begin{equation}
{\rho_{dark}\over \rho_r}\sim 10^{-8}\lambda^4{V_i\over V_F}
\end{equation}
If, on the other hand, $\rho_{dark}>V_F$ and $H\gg V_F''{}^{1/2}$, then the
field evolves very slowly along the flat direction. One of two things can
happen. The expansion rate $H$ can fall below $V_F''{}^{1/2}$
whilst $\rho_{dark}> V_F$. The field will then oscillate about the minimum of
the potential and reheat to give a radiation energy density
$\rho_r<\rho_{dark}$. The second possibility is that $\rho_{dark}$ falls below
$V_F$ whilst $H> V_F''{}^{1/2}$. In this case, we can have reheating with
$\rho_r>\rho_{dark}$ and one has to analyse the combination of
dark reheating with real reheating for the particular model to obtain the
ratios of dark radiation to real radiation.


\bibliography{paper.bib}

\end{document}